\documentclass{aa}  

\usepackage[normalem]{ulem}
\usepackage{graphicx}
\usepackage{txfonts}
\usepackage[dvipsnames]{xcolor}
\usepackage[figuresright]{rotating}
\usepackage{hyperref}
\hypersetup{
    colorlinks=true,
    allcolors=blue
    }
\usepackage{booktabs}
\begin{document}

\title{Small-scale vortical motions in cool stellar atmospheres}

\author{
      J.~R.~Canivete Cuissa\inst{1}
      \and
      F.~Riva\inst{1,2}
      \and
      O.~Steiner\inst{1,3}
}

\institute{
        Istituto ricerche solari Aldo e Cele Daccò (IRSOL), Faculty of Informatics, Università della Svizzera italiana, CH-6605 Locarno, Switzerland
     \and
        Euler Institute, Universit\`a della Svizzera italiana (USI), CH-6900 Lugano, Switzerland
     \and 
        Kiepenheuer-Institut f\"ur Sonnenphysik (KIS), 
        Sch\"oneckstrasse 6, 79104 Freiburg i.Br., Germany
}

\date{Received xxx; accepted yyy}

\abstract
{Small-scale vortices in the solar atmosphere have received considerable attention in recent years. These events  are considered potential conduits for the exchange of energy and mass between the solar atmospheric layers from the convective surface to the corona. Similar events may occur in the atmospheres of other stars and play a role in energy transfer within their atmospheres.}
{Our aim is to study the presence and properties of small-scale swirls in numerical simulations of the atmospheres of cool main-sequence stars. Our particular focus is on understanding the variations in these properties for different stellar types and their sensitivity to the surface magnetic field. Furthermore, we aim to investigate the role of these events in the energy transport within the simulated atmospheres. }
{We analyze three-dimensional, radiative-magnetohydrodynamic, box-in-a-star, numerical simulations of four main-sequence stars of spectral types K8V, K2V, G2V, and F5V. These simulations include a surface small-scale dynamo responsible for amplifying an initially weak magnetic field. Thus, we can study models characterized by very weak, or, magnetic fields in near equipartition. To identify small-scale vortices in horizontal layers of the simulations, we employ the automated algorithm SWIRL. }
{Small-scale swirls are abundant in the simulated atmospheres of all the investigated cool stars. The characteristics of these events appear to be influenced by the main properties of the stellar models and by the strength of the surface magnetic field. In addition, we identify signatures of torsional Alfv\'enic pulses associated with these swirls, which are responsible for a significant vertical Poynting flux in the simulated stellar photospheres. Notably, this flux is particularly significant in the K8V model, suggesting a possible link to the enhanced basal \ion{Ca}{ii} H and K fluxes observed in the range of $B-V$ color index $1.1 \leq B - V \leq 1.4$. Finally, we present a simple analytical model, along with an accompanying scaling relation, to explain a peculiar result of the statistical analysis that the rotational period of surface vortices increases with the effective temperature of the stellar model.  }
{Our study shows that small-scale vortical motions are not unique to the solar atmosphere and may be relevant in explaining the chromospheric activity properties of main sequence cool dwarf stars.}


\maketitle
   
%
%

\section{Introduction}
\label{sec:introduction}

Given the ubiquitous presence of small-scale swirls in the solar atmosphere \citep[see][for a review]{2023SSRv..219....1T}, it is reasonable to expect similar vortical flows in other stellar atmospheres. However, the unique properties of individual stars, such as the gravitational acceleration at the stellar surface, the effective temperature, and the mass density, may influence the properties of these small-scale features. 

In addition, magnetic fields play a crucial role in the generation and dynamics of swirls in the solar atmosphere as was demonstrated, e.g., with simulations by \citet{2020A&A...639A.118C} or \citet{2021A&A...649A.121B}. This suggests that the surface magnetic field of other stars are likewise important for stellar swirls \citep{2013AN....334..137W}. Specifically, magnetic fields appear to be tightly related to the transport of vortical motion and Poynting flux from the photosphere to the outer solar atmosphere within so called magnetic tornadoes, which possibly contribute to the heating of the chromosphere or corona \citep{2012Natur.486..505W}. 

The precise mechanism driving the upward transport of energy within magnetic tornadoes remains a topic of debate, with torsional Alfvénic pulses\footnote{The term Alfv\'enic (instead of Alfv\'en) is used to remind that these pulses or waves are not purely but mainly of Alfv\'enic nature. They may also exhibit a compressive component.} identified as the most probable candidate \citep{2019NatCo..10.3504L, 2021A&A...649A.121B}. In this scenario, small-scale swirls are the plasma counterparts to twist in the field lines of the vertically directed photospheric magnetic flux concentrations. These perturbations, originating in the surface layers of the atmosphere, can propagate upward along the flux tube into the chromosphere as pulses of torsional Alfvén waves. Moreover, numerical simulations indicate that about $80\,\%$ of the swirls in the photosphere are consistent with this scenario \citep[][]{2024A&A...inpress..C}{}{}.

Of particular interest is the possible contribution of magnetic tornadoes to the chromospheric basal flux as a function of spectral type \citep{2018A&A...616A.108B}. Observations of main-sequence stellar populations reveal that stars with $1.1 \leq B - V \leq 1.4$ exhibit a significantly enhanced chromospheric basal flux, the cause of which is not yet fully understood. Therefore, we aim to investigate the presence and properties of small-scale vortices in the simulated atmospheres of different main sequence stellar models, and their role in the transport of energy and mass through the different layers. 

For this purpose, three-dimensional radiative-magneto\-hydro\-dynamic (R-MHD) box-in-a-star numerical models have been analyzed, specifically four main-sequence stellar models of spectral types K8V, K2V, G2V, and F5V. Originally, these simulations have been performed to study the surface small-scale dynamo (SSD) in cool main-sequence stars by \citet{2024A&A...inpress..R}. Thus, with these models at hand, it became possible to investigate the properties of vortices in the presence of a magnetic field that self-consistently emerges through SSD action. We carry out a statistical analysis of the fundamental properties of the vortices during both the kinematic and saturation phases of the SSD in different heights within the simulated atmosphere, and across different stellar models.

The paper is organized as follows: Sect.\,\ref{sec:methods} describes the simulations and the vortex identification algorithm used in this study. In Sect.\,\ref{sec:results_and_discussion}, we present results on the qualitative and quantitative properties of the vortices in the four stellar models. In addition, we introduce a simple analytical model for photospheric vortices that replicates one of the results of our statistical analysis. Finally, Sect.\,\ref{sec:conclusions} provides a comprehensive summary and the conclusions.
%
%

\section{Methods}
\label{sec:methods}


\subsection{Numerical simulations}
\label{subsec:numerical_simulations}

The three-dimensional R-MHD simulations that serve as the basis of the present investigation have been carried out with the CO5BOLD code \citep{2012JCoPh.231..919F} in the so called box-in-a-star setup. This corresponds to simulating a small partial volume near the optical surface of a star, encompassing the layers where energy transport changes from predominantly convective to purely radiative. A constant and vertical gravitational acceleration is applied. The simulations use the long-characteristic radiative transfer module of CO5BOLD with Rosseland mean opacities and the diffusion approximation in the regime of high optical depths ($\log\tau_{\rm R} \gtrsim 2$). Details on the numerical setup can be found in \citet{2024A&A...inpress..R}.

\begin{table}
    \renewcommand\arraystretch{1.2}
    \centering
    \caption{Parameters describing the four simulated stellar models. See text in Sect.\,\ref{subsec:numerical_simulations} for more details.}
    \begin{tabular}{rcccc}
    \hline\hline 
    Spectral Type & K8V & K2V & G2V & F5V \\
    \hline
    $T_{\rm eff}\,[{\rm K}]$ & 4005 & 5000 & 5766 & 6506 \\
    $\log{(g)}$  & 4.66 & 4.59 & 4.44 & 4.24 \\ 
    $H_{\rm p}(z=0)\,[{\rm km}]$ & 60 & 100 & 182 & 348 \\
    \hline
    $L_x, L_y\, [{\rm km}]$ & 2200 & 3500 & 6000 & 15000 \\
    $L_z \,[{\rm km}]$ & 968 & 1540 & 3072 & 7800 \\
    Below $z=0$ [km] & 748 & 1190 & 2412 & 6150 \\
    Above $z=0$ [km] & 220 &  350 &  660 & 1650 \\
    $\Delta x, \, \Delta y, \, \Delta z\,[{\rm km}]$ & 2.93 & 4.67 & 8.00 & 20.00 \\
    $L_{\rm gran}\, [{\rm km}]$ & 354 & 609 & 990 & 2490 \\
    \hline
    $N_{\rm kin}$ & 27 & 20 & 18 & 20\\
    $N_{\rm sat}$ & 20 & 32 & 36 & 34 \\
    \hline
    \end{tabular}
    \tablefoot{$T_{\rm eff}$, $L_{\rm gran}$, and $H_P$ are evaluated as time averages during the kinematic phase of the simulations. For more details, the reader can refer to \citet{2024A&A...inpress..R}.}
    \label{tab:sim_info}
\end{table}

The stellar models are primarily distinguished by their effective temperature, $T_{\rm eff}$, gravitational acceleration, $g$, and the chemical composition (metallicity), which here is taken to be the solar one.  The main parameters $g$ and the average $T_{\rm eff}$ of the models are provided in the top set of rows in Tab.\,\ref{tab:sim_info} together with the average pressure scale height at the surface of the model, $H_{\rm p}(z=0)$. 
Here, surface denotes the space and time averaged optical depth surface $\tau_{\rm R} = 1$, which is also taken as the origin of the geometrical $z$-scale, $z=0$. The surface gravity $g$ was chosen to be that of a dwarf star placed on the main sequence.

The middle set of rows of Tab.\,\ref{tab:sim_info} specifies the horizontal and vertical sizes of the simulated domains, $L_x$, $L_y$, and $L_z$, along with the grid spacing, indicated by $\Delta x$, $\Delta y$, and $\Delta z$. The vertical extension, $L_z$, is chosen such that the bottom boundary is at a depth where the entropy of the downflowing plasma approaches the essentially constant entropy of the upflowing plasma \citep[see,][Fig.\,11]{2018A&A...614A..78S}, whereas the top boundary is located well above the entropy minimum in the photosphere. The height extents of the computational domains, below and above the average Rosseland optical depth unity, are also displayed in this section of the table. Also given is the average size of a granule, $L_{\rm gran}$, wich is computed as the integral length scale of the vertical velocity spectrum according to Eq.\,(5) of \citet{2024A&A...inpress..R}. 
The horizontal extent is specifically chosen to encompass approximately six granular scales within the computational domain. Vertically, the simulation domains extend from the near surface layers of the convective zone to beyond the entropy minimum of the photosphere, corresponding to approximately the top layers of the photosphere. 

Pre-existing and relaxed purely hydrodynamic simulations were seeded with a vertical magnetic field of $B_z = 1\,{\rm mG}$. These simulations were then further evolved and underwent a kinematic phase in which the seed magnetic field was exponentially amplified by the action of a sub-surface SSD until it reached saturation. The bottom set of rows indicates the number of analyzed snapshots during the kinematic and saturated phases,  $N_{\rm kin}$ and $N_{\rm sat}$, respectively. The selection of snapshots for the kinematic and saturated phases is based on the mean ratio between kinetic and magnetic energy densities in the convection zone, $\langle E_{\rm kin}/E_{\rm mag} \rangle_{\rm conv}$. We consider a time instance to be in the kinematic phase when $\langle E_{\rm kin}/E_{\rm mag} \rangle_{\rm conv} < 2.5 \cdot 10^{-3}$, while saturation\footnote{For the four simulations, the maximum values of $\langle E_{\rm kin}/E_{\rm mag} \rangle_{\rm conv}$ are around 0.10. For further details, the reader can refer to \citet{2024A&A...inpress..R}.} is defined when $\langle E_{\rm kin}/E_{\rm mag} \rangle_{\rm conv} > 2.5 \cdot 10^{-2}$.


\subsection{Identification algorithm}
\label{subsec:swirls_identification}

We employ the method proposed by \citet{2022A&A...668A.118C,2024A&A...inpress..C} to identify swirls in numerical simulations. This method accurately estimates the coordinates of the center of rotation, that is, the estimated vortex centers (EVC), for each fluid particle (grid cell) that exhibits some degree of curvature in the instantaneous velocity field. 
Since vortices can be viewed as groups of fluid particles coherently rotating about a common axis of rotation \citep{1979rdte.book..309L}, clusters of EVCs form in the neighborhood of the core of a vortex structure. Therefore, the method uses clusters of EVCs to identify vortices.
In this respect, the EVC method can be seen as a more accurate version of the curvature center method proposed by \citet{Sadar99}, since it uses the velocity field and its derivatives to extract more detailed information about the curvature of the flow. 

The accuracy of the method hinges on an accurate estimate of the radius of curvature and radial direction of the local flow. This requires the measurement of the rigid-body rotational component of the flow. Traditional methods have used mathematical quantities such as vorticity or swirling strength \citep{1999JFM...387..353Z} for this purpose, but \citet{2022A&A...668A.118C} showed that the optimal quantity is the Rortex criterion $R$ \citep{2018JFM...849..312T, 2019JHyDy..31..464W}. This advanced mathematical criterion measures alone the rigid body rotational component of a rotating flow and can be computed as,
\begin{equation}
    R = \boldsymbol{\omega}\cdot\boldsymbol{u}_{\rm r} - \sqrt{\left(\boldsymbol{\omega}\cdot\boldsymbol{u}_{\rm r}\right)^2 - \lambda^2} \, , \label{eq:rortex}
\end{equation}
where $\boldsymbol{\omega}$ is the vorticity vector, $\boldsymbol{u}_{\rm r}$ is the normalized, real eigenvector of the Jacobian of the velocity vector, and $\lambda$ is the swirling strength criterion, which formally is twice the imaginary part of the complex conjugated eigenvalues of the Jacobian of the velocity vector. For more details on these quantities, the reader can refer to \citet{2020A&A...639A.118C,2022A&A...668A.118C}. 

Clusters of EVCs can in principle be identified by eye. However, \citet{2022A&A...668A.118C} adapted the algorithm named clustering by fast search and find of density peaks \citep[CFSFDP,][]{2014Sci...344.1492R} to automate the process and reduce human bias in the identification. In addition, the proposed algorithm includes a cleaning procedure to remove misidentifications caused by noise or coherent but non-spiraling curved flows. 

The algorithm outputs a list of identified vortices, each characterized by its center coordinates, the region it occupies in a horizontal section (in grid-cells), and its rotation direction (clockwise or counterclockwise). To quantify the effective radius of the swirl, $r_{\rm eff}$, the following formula is employed,
\begin{equation}
    r_{\rm eff} = \sqrt{\frac{N_{\rm c}}{\pi}} \Delta x \, , \label{eq:r_eff}
\end{equation}
where $N_{\rm c}$ represents the number of grid cells occupied by the swirl and $\Delta x$ is the grid spacing. Additionally, meaningful properties can be computed by combining this information with the values of physical variables on the swirl surface. For instance, the effective period of rotation, $P_{\rm eff}$, can be estimated using,
\begin{equation}
    P_{\rm eff} = \frac{4\pi}{\langle R_z \rangle_{\rm swirl}} \, , \label{eq:P_eff}
\end{equation}
where $\langle\,\cdot\,\rangle_{\rm swirl}$ denotes the spatial average over the swirl surface, and $R_z$ represents the vertical component of the Rortex vector, $\boldsymbol{R}$. Another noteworthy property is the effective plasma rotational velocity, $v_{\phi,\,{\rm eff}}$, determined by,
\begin{equation}
    v_{\phi,\,{\rm eff}} = \frac{2 \pi r_{\rm eff}}{P_{\rm eff}} \, . \label{eq:v_phi_eff}
\end{equation}

A Python implementation of the resulting algorithm is open source on GitHub\footnote{\url{https://github.com/jcanivete/swirl}} and named SWirl Identification by Rotation-centers Localization \citep[SWIRL,][]{SWIRL}. For more details on the method, the clustering algorithm, and the test cases, the reader can refer to \citet{2022A&A...668A.118C}.

%
%

\section{Results and discussion}
\label{sec:results_and_discussion}

In this section, we present and discuss the results of our analysis of the properties of the swirls in the simulated stellar atmospheres. First, we qualitatively validate the vortex identification algorithm, SWIRL, by comparing its output with emerging bolometric intensity maps and horizontal velocity fields. We then perform a comprehensive statistical analysis of the swirls properties, exploring their tight relationship with magnetic fields and study their energetics. Finally, we present a simple model of a photospheric vortex that supports a surprising result revealed by the statistical analysis.


\subsection{Qualitative validation of SWIRL}
\label{subsec:qualitative_validation_of_SWIRL}

\begin{figure*}
	\centering
	\resizebox{\hsize}{!}{\includegraphics{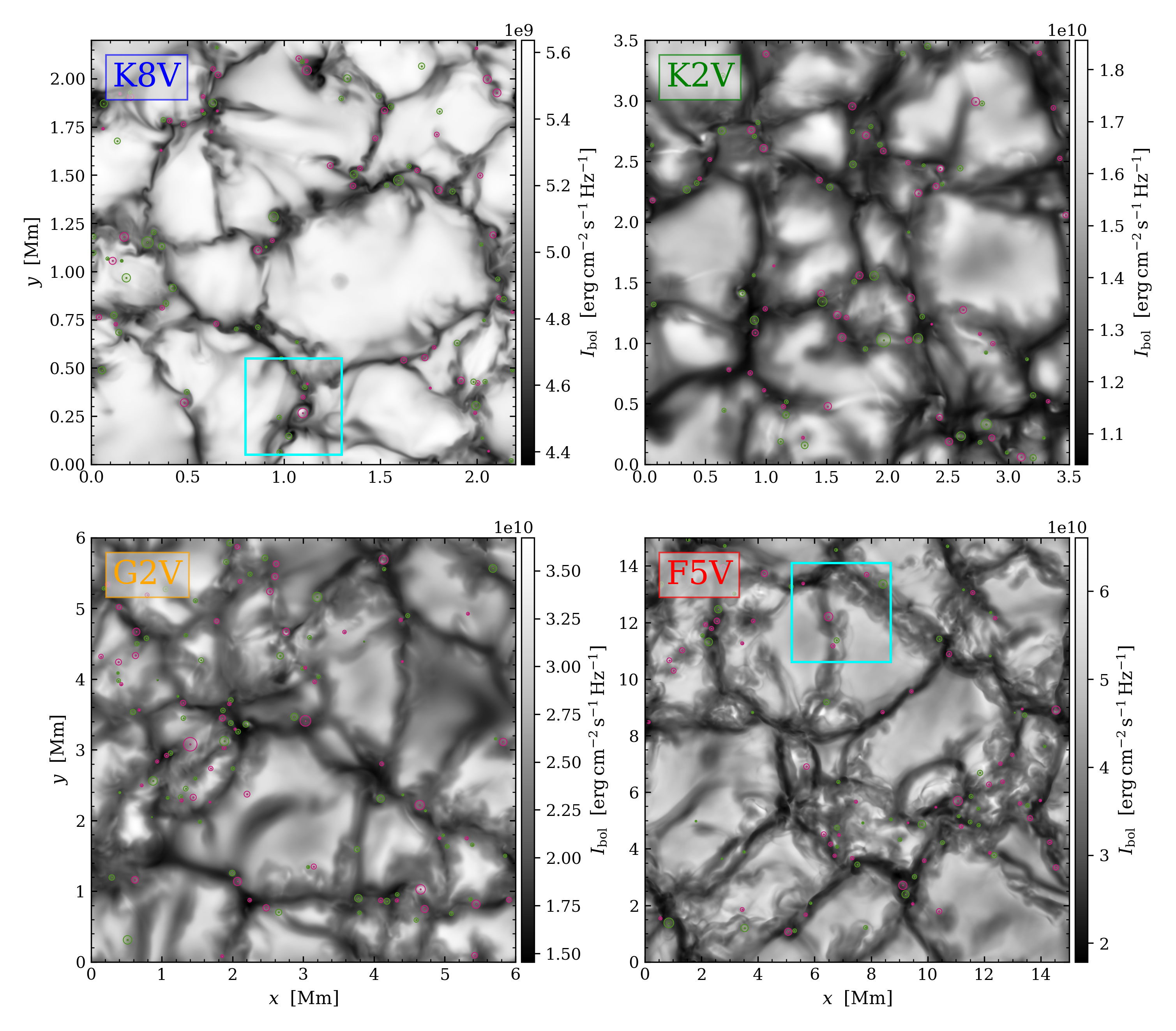}}
	\caption{Emerging bolometric intensity, $I_{\rm bol}$, for the K8V, K2V, G2V, and F5V models during the kinematic phase. Clockwise and counterclockwise vortices identified by the SWIRL algorithm are indicated by pink and green disks, respectively. The blue squares in the K8V and F5V panels mark the boundaries of the zoomed plots shown in Fig. \ref{fig:visualization_zoomin_Ic}.}
	\label{fig:visualization_Ic}
\end{figure*}
\begin{figure*}
	\centering
	\resizebox{\hsize}{!}{\includegraphics{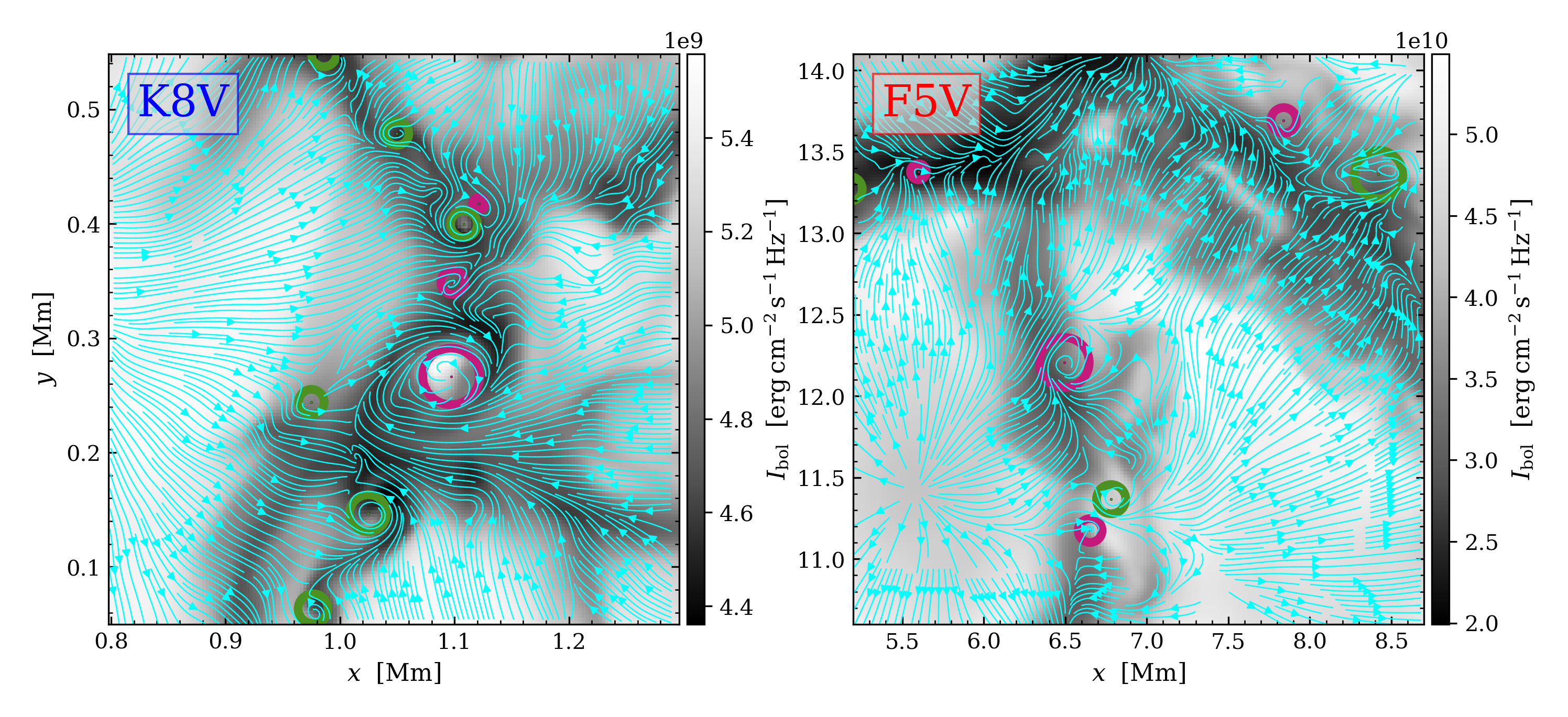}}
	\caption{Identified swirls in the areas marked in blue in Fig.\,\ref{fig:visualization_Ic} for the K8V and F5V models. Clockwise and counterclockwise vortices identified by the SWIRL algorithm are indicated by pink and green disks, respectively. The horizontal component of the velocity field is represented by blue instantaneous streamlines.}
	\label{fig:visualization_zoomin_Ic}
\end{figure*}

Figure \ref{fig:visualization_Ic} shows the emerging bolometric intensity, $I_{\rm bol}$, at a specific time during the kinematic phase for the four stellar models. At this time, the magnetic field remains relatively weak and exerts only a feeble influence on the flow dynamics. The granulation, characterized by cells of bright (warm) upflowing plasma surrounded by darker (cooler) downflow in the intergranular space, is clearly visible in all models. In particular, the granules in the cooler stars (K8V and K2V) exhibit a smoother surface compared to the more vigorous dynamics visible in the G2V and F5V models.

From the horizontal velocity field at the surface ($z=0\,{\rm km}$) and at the corresponding specific times, the SWIRL algorithm identifies the vortices shown in the four panels of Fig. \ref{fig:visualization_Ic}. Each identified vortex is represented by a circle of size $r_{\rm eff}$, denoting the effective radius of the vortex, and color-coded according to the direction of rotation (clockwise or counterclockwise). In Fig.\,\ref{fig:visualization_zoomin_Ic}, the identified vortices, accompanied by instantaneous streamlines of the horizontal velocity field, are shown in detail in two zoomed regions for the K8V and the F5V model.

The majority of the vortices are located within intergranular lanes, consistent with the results reported for simulations \citep{2024A&A...inpress..C} and observations \citep[][]{2008ApJ...687L.131B}{}{}  of the solar photosphere . This behavior is expected due to the conservation of angular momentum in intergranular downflows \citep{1985SoPh..100..209N, 2014PASJ...66S..10W}. An examination of the instantaneous streamlines in the zoomed plots of Fig.\,\ref{fig:visualization_zoomin_Ic} shows that the SWIRL algorithm consistently identifies vortical patterns in the velocity field. This is true regardless of whether the flow appears more laminar-like, as in the K8V model, or more vigorous and highly turbulent, as in the F5V model.

In Fig.\,\ref{fig:visualization_Ic} it can be seen that certain swirls, though not all, exhibit increased radiative intensity compared to their surroundings. An illustration of such a case is in the left panel of the Fig.\,\ref{fig:visualization_zoomin_Ic} located at $(x,y) = (1.10,0.25)\,{\rm Mm}$. For comparison, in Appendix\,\ref{app:surface_distribution_density} we provide plots of the plasma mass density $\rho$ at $z=0\,{\rm km}$ during the same time instances as shown in Figs.\,\ref{fig:visualization_Ic} and \ref{fig:visualization_zoomin_Ic}. We notice that areas of enhanced bolometric intensity in the integranular space that are associated with swirling motions exhibit significantly reduced mass densities, as is particularly well visible in Fig.\,\ref{fig:visualization_zoomed_rho}. Partial evacuation and enhanced radiative intensity in connection with swirling motions are the characteristic ingredients of so called non-magnetic bright points. Non-magnetic bright points occurring in magnetic field free simulations of the solar atmosphere have been investigated by \citet{2016A&A...596A..43C}. Here, they frequently occur under weak magnetic conditions in the kinematic phase of all stellar models. 

In the saturation phase of the SSD, the basic characteristics and distribution of the swirls in the bolometric intensity maps, and the reliability of their detection with the SWIRL algorithm carries over from the kinematic phase. Therefore, we have chosen not to show further similar plots here. The primary difference lies in the increased prevalence of bright features within intergranular regions, attributed to magnetic flux concentrations \citep[see,][Fig.\,2]{2024A&A...inpress..R}{}{}, that appear to correlate with the presence of vortices. This aspect is quantitatively investigated in Sect.\,\ref{subsubsec:surface_distrutions}.


\subsection{Statistical Analysis}
\label{subsec:statistical_analysis}

\begin{table*}
    \renewcommand\arraystretch{1.2}
    \centering
    \caption{Location of the selected horizontal planes for the statistical analysis.}
    \vspace{0.2cm}
    \begin{tabular}{rccccc}
    \hline\hline 
    Plane & $\ln(p_0/p)$ & $z_{K8V}\,[{\rm km}]$ & $z_{K2V}\,[{\rm km}]$ & $z_{G2V}\,[{\rm km}]$ & $z_{F5V}\,[{\rm km}]$ \\
    \hline
    Convection zone & -0.5 & -31 & -57 & -110 & -206  \\
    Surface & 0.0 & 0 & -1 & -2 & -6 \\
    Middle photosphere & 2.0 & 101 & 153 & 270 & 475 \\
    High photosphere & 4.0 & 183 & 284 & 486 & 835 \\
    \hline
    \end{tabular}
    \tablefoot{The values of $z$ for each model correspond to the plane that best approximates the mean value of $\ln(p_0/p)$ relative to the mean $\tau_{\rm R} = 1$ surface.}
    \label{tab:planes}
\end{table*}

In this subsection, we examine the characteristics of the vortices identified by the SWIRL algorithm in both the kinematic and saturated phases. The number of snapshots analyzed in each regime for each stellar model is detailed in the bottom set of rows of Tab.\,\ref{tab:sim_info}.

Our analysis refers to four different heights in the simulated atmospheres, roughly corresponding to the surface layer of the convection zone, the stellar surface, the middle photosphere, and the high photosphere. The specific location is determined by the quantity $\ln(p_0/p)$, where $p_0$ is the horizontally averaged pressure at the mean optical depth $\tau_{\rm R}=1$ ($z\approx 0\,{\rm km}$), and $p$ is the horizontally averaged pressure at the targeted plane. Table~\ref{tab:planes} gives the assumed values of $\ln(p_0/p)$ along with the corresponding heights $z$ in the simulation domains of the four models.

\subsubsection{Surface distributions}
\label{subsubsec:surface_distrutions}
\begin{figure}
	\centering
	\resizebox{\hsize}{!}{\includegraphics{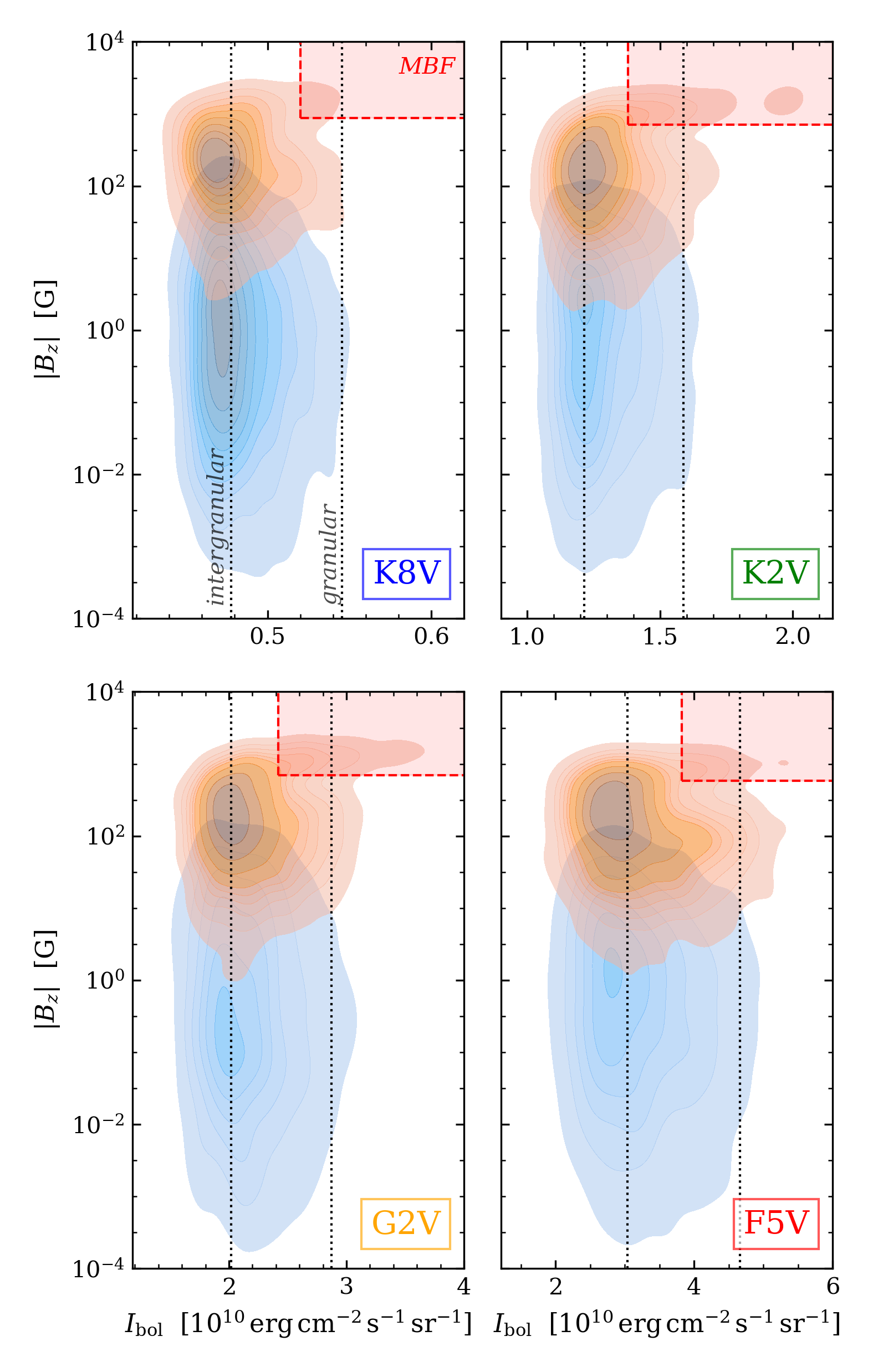}}
	\caption{Bivariate distribution of the bolometric intensity $I_{\rm bol}$ and the absolute vertical magnetic field $|B_z|$ in the center of the swirls identified on the surface layer at $z=0\,{\rm km}$ of the four stellar models. The distributions during the kinematic and saturation phases are shown in blue and orange, respectively. The vertical black dotted lines indicate the average bolometric intensity in the intergranular (lower value) and granular (higher value) regions. The red shaded area indicates the identification range of magnetic bright feature (MBF) as defined by \citet{2024A&A...inpress..R}.}
	\label{fig:stats_Bz_vs_Ic}
\end{figure}

To further investigate the distribution of vortices with respect to the granular configuration, Fig.\,\ref{fig:stats_Bz_vs_Ic} shows the bivariate distribution of the bolometric intensity $I_{\rm bol}$ and the absolute vertical magnetic field $|B_z|$ for all the swirls identified at $z=0\,{\rm km}$. The values of $I_{\rm bol}$ and $|B_z|$ are taken at the center of each swirl. Moreover, we distinguish swirls identified during the kinematic phase (blue) from swirls in the saturated phase (orange) of the sub-surface SSD.

The vertical dashed lines in Fig.\,\ref{fig:stats_Bz_vs_Ic} denote the average bolometric intensity in the intergranular and granular regions for the four stellar models. These regions are defined based on the vertical velocity of the plasma, $v_z$, at the surface ($z=0\,{\rm km}$). Specifically, grid cells within the intergranular and granular regions satisfy the conditions $v_z < -\sigma_{v,z}$ and $v_z > \sigma_{v,z}$, respectively, where $\sigma_{v,z}$ is the standard deviation of the distribution of $v_z$ at $z=0\,{\rm km}$.

During the kinematic phase (blue), the majority of the swirls in all models are located in regions of low bolometric intensity, corresponding to intergranular lanes. This finding confirms the visual impression conveyed by Fig.\,\ref{fig:visualization_Ic}. The high end tails of the four distributions extend into the range of average granular bolometric intensity, indicating the presence of swirls associated with non-magnetic bright points (see left panel of Fig.\,\ref{fig:visualization_zoomin_Ic}) or within granules. As for the distribution in the vertical magnetic field, it spans several orders of magnitude ($10^{-4}\,{\rm G} \lesssim |B_z| \lesssim 10^2\,{\rm G}$) as the magnetic field is amplified by the sub-surface SSD, starting from an initial value of $1\,{\rm mG}$. Despite the large range, the distribution appears symmetrical in all four models. We conclude that the location of the swirls is essentially independent of the magnetic fields during this phase.

In the saturated phase (red), the distributions in $I_{\rm bol}$ remain centered around the intergranular bolometric intensities for the four models. At the same time, the distributions of $|B_z|$ become narrower and shift to higher values ($1\,{\rm G} \lesssim |B_z| \lesssim 10^4 \,{\rm G}$) due to stronger magnetic fields in the saturation phase.

A notable feature emerges in all panels: the bivariate distributions exhibit a distinct presence of swirls in regions of high $I_{\rm bol}$ and near $\max(|B_z|)$, particularly evident in the K2V and G2V models. This observation strongly suggests the presence of swirls associated with magnetic bright features (MBFs), a phenomenon previously documented in both solar observations \citep[see, e.g.,][]{2008ApJ...687L.131B} and simulations \citep[see, e.g.,][]{2011A&A...533A.126M}. Notably, the red boxes in the top-right corners of the four panels indicate the identification range for MBFs, as defined in \citet[][Appendix B]{2024A&A...inpress..R}. This indicates that, for a sufficiently strong surface magnetic field, swirls are found in correspondence with MBFs across different stellar models. 


\subsubsection{Physical properties}
\label{subsubsec:swirl_physical_properties}

\begin{figure}
    \centering
    \resizebox{\hsize}{!}{\includegraphics{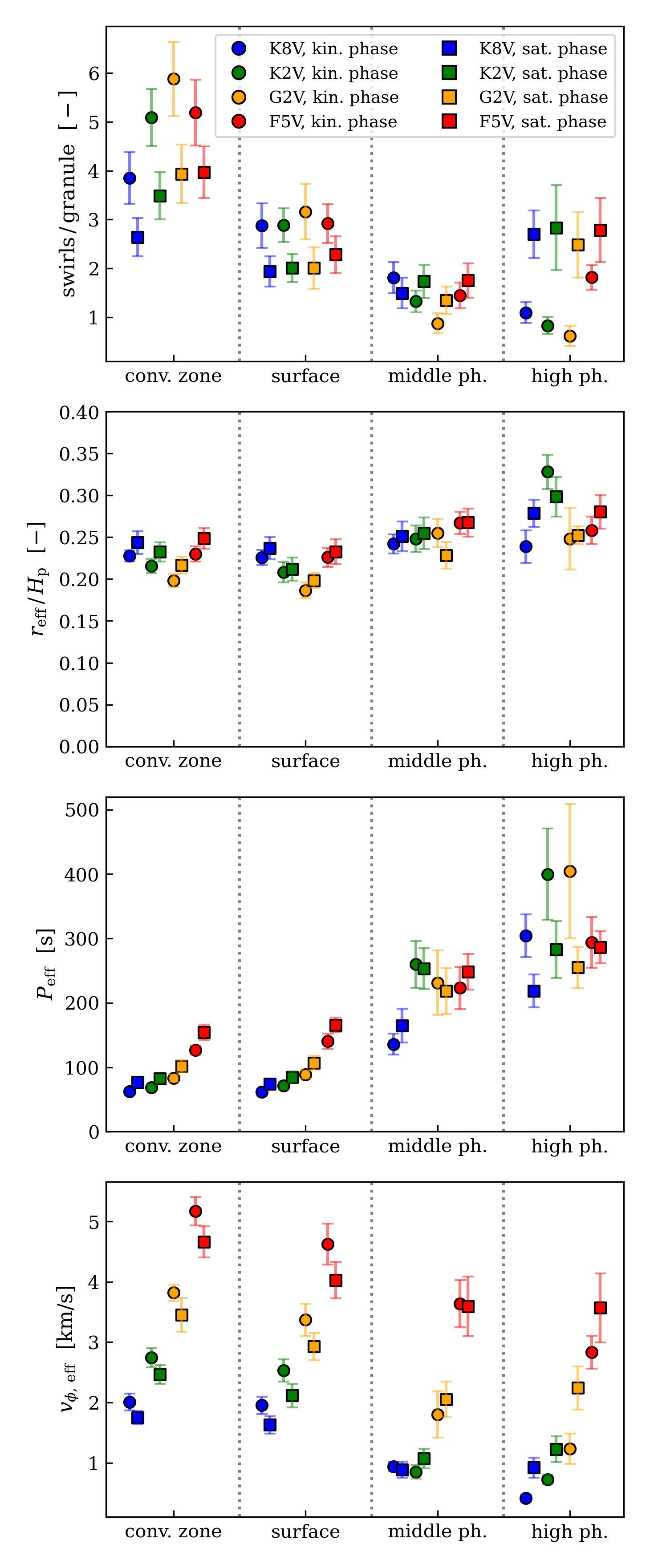}}
    \caption{Properties of the swirls during the kinematic and saturation phases of the SSD and at different heights in the atmosphere of the four stellar models. The average number of swirls per granule in the simulation domain (\textit{top}), the average ratio between radius and pressure scale height, $r_{\rm eff}/H_{\rm p}$ (\textit{middle-top}), rotational period, $P_{\rm eff}$ (\textit{middle-bottom}), and plasma rotational velocity $v_{\phi,\,{\rm eff}}$  (\textit{bottom}), are shown. The heights of the cross sections labelled as convection zone, surface, middle photosphere, and higher photosphere are given in Tab.\,\ref{tab:planes} for the four stellar models.}
    \label{fig:stats_swirl_properties}
\end{figure}

Next, we investigate the main physical characteristics of swirls across the four stellar models, the four heights in the simulation domain, and the two phases of the sub-surface SSD. The physical properties we are interested in are: the average number of swirls per granule, the average ratio between the effective radius and the pressure scale height at $z=0\,{\rm km}$, $r_{\rm eff}/H_{\rm p}$, the average effective period of rotation $P_{\rm eff}$, and the average rotational velocity $v_{\phi,\,{\rm eff}}$. The approximate size of the granules and pressure scale heights at $z=0\,{\rm km}$ for the four models are given in Tab.\,\ref{tab:sim_info}, while for computing $r_{\rm eff}$, $P_{\rm eff}$, and $v_{\phi,\,{\rm eff}}$ we employ Eqs.\,(\ref{eq:r_eff}), (\ref{eq:P_eff}), and (\ref{eq:v_phi_eff}), respectively. 

The results of the statistical analysis are shown in Fig.\,\ref{fig:stats_swirl_properties}. Focusing on the kinematic phase, the top panel of the figure highlights that the convection zone has the highest number of vortices per granule in all four models. This number is computed as the total number of vortices in the corresponding horizontal plane divided by the area of this plane, $L_x\cdot L_y$, times the approximate area of a granule $L_{\rm gran}^2$. As we ascend through the atmosphere, the number of vortices per granule gradually decreases. These results are consistent with simulations of the solar atmosphere \citep[see][]{2012A&A...541A..68M, 2024A&A...inpress..C}. The prevalence of highly turbulent convective flows in the sub-surface layers of cool stars naturally leads to the formation of vortex motions that explain the observed trend. Interestingly, within the convection zone, the G2V model stands out as having the highest number of vortices among the different models.

Conversely, the F5V model has the highest number of vortices in the high photosphere. This can be attributed to the energetic nature of the flows observed in the F5V model, which leads us to expect the presence of shocks already in the photosphere. These shocks could serve as a source of vortex motions in the upper atmospheres, especially in the virtual absence of magnetic fields \citep{2012A&A...541A..68M}.

During the saturation phase, there are significant changes in the number of vortices in the convection zone and in the high photosphere. In the convection zone, the presence of strong magnetic fields leads to a significant suppression of vortex formation. This suppression results from the ``stiffening'' effect of magnetic fields on the plasma as it approaches equipartition -- a phenomenon known as magnetic quenching \citep[see, e.g.,][]{2005PhR...417....1B}.

In contrast, the number of vortices in the high photosphere increases significantly, by a factor of about 3 to 4, except for the F5V model where the increase is less strong. This observation underscores the central role of magnetic fields in the generation and dynamics of vortex features in the upper atmosphere. The results are consistent with simulations of the solar atmosphere performed by \citet{2012A&A...541A..68M} and \citet{2021A&A...649A.121B}. Moreover, the similar number of vortices in the high photosphere across models indicates that magnetic effects outweigh shocks in the formation of vortices within the magnetized upper photosphere. These results are in agreement with a similar conclusion presented in \citet{2020A&A...639A.118C}.

The second panel of the Fig.\,\ref{fig:stats_swirl_properties} shows the mean ratio between the effective radius of the swirl, $r_{\rm eff}$, and the pressure scale height at $z = 0\,{\rm km}$, $H_{\rm p}$. This ratio allows a more meaningful comparison of results between different models, taking into account the different scales inherent in each simulation. In stars characterized by larger granules, such as the F5V model, one would expect the swirls to be larger due to the broader scales of the flows. Therefore, by normalizing the radius by the surface pressure scale height -- an essential length scale in the models -- we obtain a normalized radius that allows a direct and fair comparison between the different models.

The results show minimal variation in the mean normalized radius of vortices across spectral type, model layers, and magnetic field strengths, with values around $r_{\rm eff}/H_{\rm p} \sim 0.25$. 
Notably, the only discernible trend indicates a subtle increase in the photosphere. This result is consistent with the statistical analysis of the simulated solar atmosphere presented in \citet{2024A&A...inpress..C}. The result suggests an intrinsic relationship between the average size of the swirls and the typical length scale of the simulated atmosphere. Given that the pressure scale height is related to the characteristic size of granular flows (see Tab.\,\ref{tab:sim_info}), this observed relationship seems reasonable.

The bottom two panels of Fig.\,\ref{fig:stats_swirl_properties} show the average rotational period, $P_{\rm eff}$, and the rotational velocity, $v_{\phi,\,{\rm eff}}$, of the swirls. An interesting observation is the correlation between the rotational velocity and the effective temperature of the stellar model at all heights. This correlation is expected, since higher effective temperatures generally correspond to higher plasma velocities. However, in contrast to this, we also observe the counterintuitive result that the rotational period also increases. Obviously, the increasing length scale with increasing effective temperature compensates for the higher plasma velocity. Predicting which effect dominates is not trivial. We further explore this intriguing finding with a simple analytical model in Sect.\,\ref{subsec:temperature_period_relation}.

In the saturation phase of the SSD, a noticeable trend emerges: swirls in the sub-surface and surface layers of the simulated atmospheres show slower rotation than in the kinematic phase. This phenomenon can be attributed to the conversion of turbulent kinetic energy into magnetic energy by the SSD, resulting in a deceleration of the rotation of vortices generated by turbulent convective plasma. However, an interesting contrast occurs in the high photosphere, where vortices tend to rotate faster in the saturation phase of the SSD.


\subsubsection{Alfvénic properties}
\label{subsubsec:alfvénic_properties}

\begin{figure*}
	\centering
	\resizebox{\hsize}{!}{\includegraphics{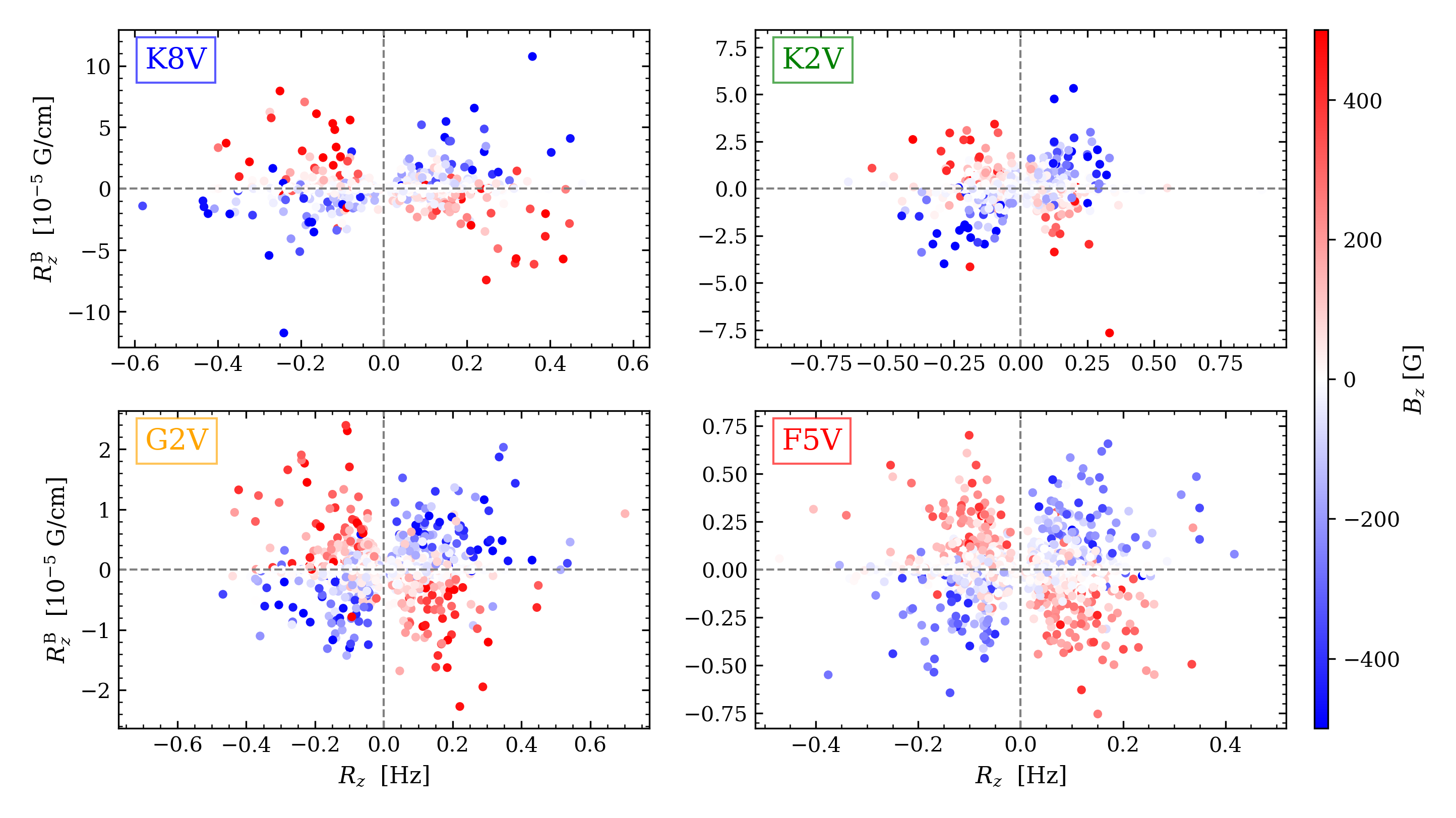}}
	\caption{Bivariate distribution of the rotational characteristics of vortices in the middle photosphere during the saturated phase of the SSD for the four stellar models. Every identified vortex is represented by a scatter point according to the Rortex criterion $R_z$ and the magnetic Rortex criterion $R_z^{\rm B}$ averaged over its area. The vertical magnetic field $B_z$ averaged over the swirl area is color-coded.}
	\label{fig:stats_alfven}
\end{figure*}
\begin{figure*}
	\centering
	\resizebox{\hsize}{!}{\includegraphics{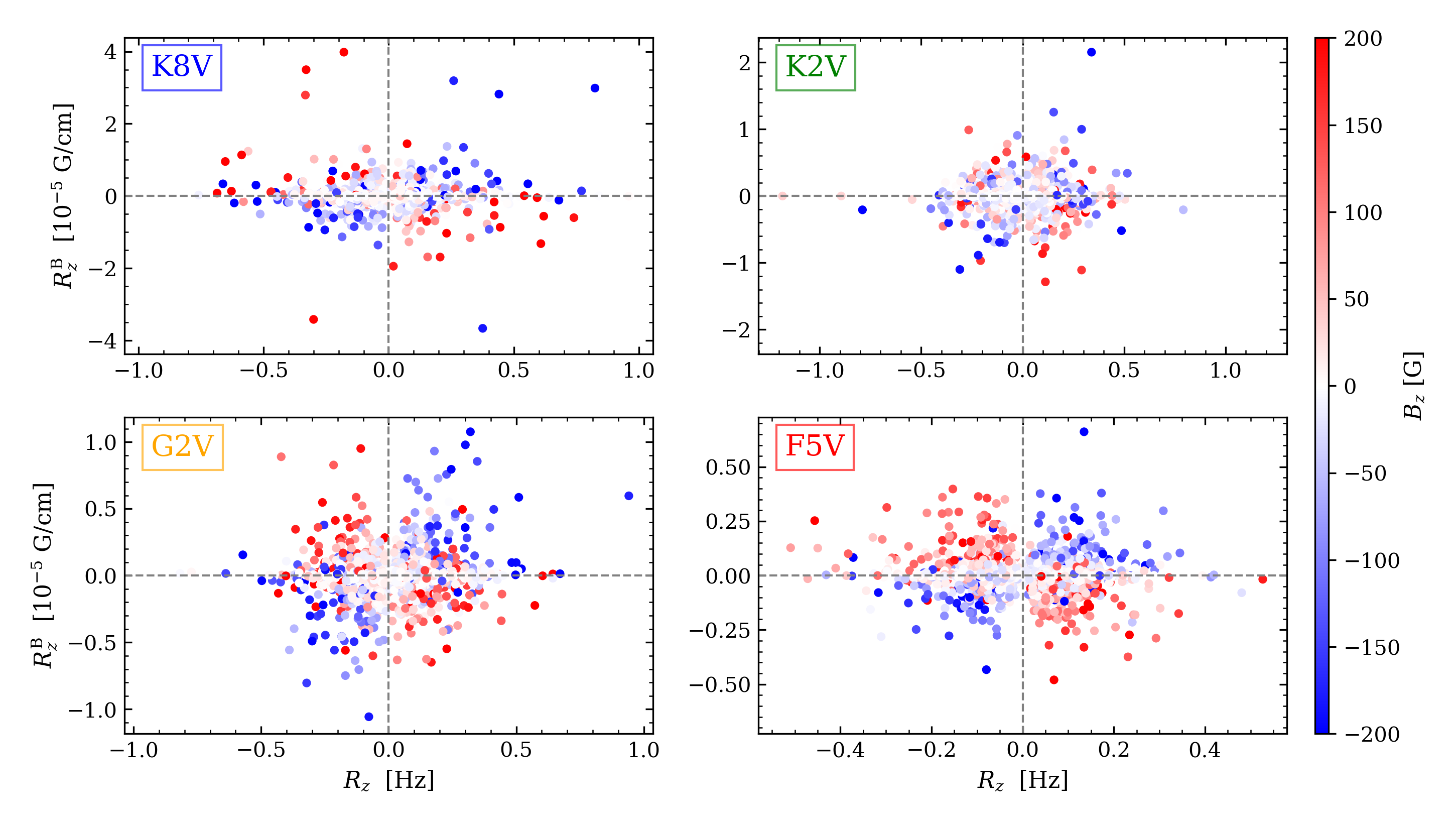}}
	\caption{Same as Fig.\,\ref{fig:stats_alfven} but for the high photosphere.}
	\label{fig:stats_alfven_highphsphere}
\end{figure*}

Sections \ref{subsubsec:surface_distrutions} and \ref{subsubsec:swirl_physical_properties} underscore the fundamental role played by magnetic fields in shaping the properties of the swirls in the four models of stellar photospheres. Therefore, the goal of this subsection is to investigate this relationship in more detail. Of particular interest is the possible association of swirls in magnetized regions with torsional Alfvénic pulses \citep[see, e.g.,][]{2012Natur.486..505W, 2019NatCo..10.3504L, 2021A&A...649A.121B}. These pulses provide a mechanism for energy transport from the surface atmospheric layer to the chromosphere and above.

\citet{2024A&A...inpress..C} conducted a thorough analysis of the swirl properties identified in simulations of the solar atmosphere. Their investigation focused on the imprints of torsional Alfvénic waves in atmospheric swirls, showing that about $80\,\%$ of photospheric swirls are consistent with the scenario of a torsional Alfvénic pulse. Next, we reproduce part of this analysis for the four stellar models.

We begin by considering an incompressible plasma in magneto-hydrostatic equilibrium within the ideal MHD approximation. Then, for a torsional Alfvén wave propagating in the vertical direction along a vertically oriented magnetic field $\boldsymbol{B} = B_z \boldsymbol{e_z}$, the dispersion relation of the wave results in,
\begin{equation}
    \boldsymbol{v}^{\prime} = - \frac{v_{\rm A}}{B_z} \boldsymbol{b}^{\prime}\,, \label{eq:alfvenwave_simple}
\end{equation}
where $\boldsymbol{v}^{\prime}$ and $\boldsymbol{b}^{\prime}$ 
denote plasma velocity and magnetic field perturbations in the horizontal plane, respectively, while $v_{\rm A} = |\boldsymbol{B}|/\sqrt{4\pi\rho} \geq 0$ 
is the local Alfvén speed. A detailed derivation can be found in \citet[][Sect. 4.3.1]{2014masu.book.....P}.

From Eq.\,(\ref{eq:alfvenwave_simple}), we deduce that the perturbations $\boldsymbol{v}^{\prime}$ and $\boldsymbol{b}^{\prime}$ are either parallel or anti-parallel, depending on the polarity of the vertical magnetic field, determined by the sign of $B_z$. Furthermore, assuming that the velocity perturbation manifests itself as a vortex, that is a rotational motion of the plasma in the horizontal plane, the magnetic field perturbation corresponds to the horizontal component of a twist in the predominantly vertically oriented magnetic field lines.

To quantify the horizontal velocity perturbations in the plasma, $\boldsymbol{v}^{\prime}$, we use the vertical component of the Rortex vector, $R_z$. Similarly, we define the magnetic Rortex vector, $\boldsymbol{R}^{\rm B}$, which measures the equivalent to the rigid-body rotational component for the torsional component of the magnetic field. The magnetic Rortex vector is defined as the Rortex vector, Eq.\,(\ref{eq:rortex}), but with the magnetic field $\boldsymbol{B}$ as the input field, resulting in,
\begin{equation}
    \boldsymbol{R}^{\rm B} = \boldsymbol{\tilde{J}}\cdot\boldsymbol{u}_{\rm r}^{\rm B} - \sqrt{\Big(\boldsymbol{\tilde{J}}\cdot\boldsymbol{u}_{\rm r}^{\rm B}\Big)^2 - \Big(\lambda^{\rm B}\Big)^2} \, , \label{eq:magnetic_rortex}
\end{equation}
where $\boldsymbol{\tilde{J}} = \boldsymbol{\nabla}\times\boldsymbol{B}$ is proportional to the current density, $\boldsymbol{u}_{\rm r}^{\rm B}$ is the normalized, real eigenvector of the Jacobian of the magnetic field, and $\lambda^{\rm B}$ is the magnetic swirling strength \citep{2021A&A...649A.121B}.

Using the vertical components of the Rortex and magnetic Rortex vectors, $R_z$ and $R_z^{\rm B}$, as proxies to quantify the perturbations associated with upward propagating torsional Alfvénic pulses, we can express Eq.\,(\ref{eq:alfvenwave_simple}) as,
\begin{equation}
    {\rm sign}{\Big( R_z R_z^{\rm B} \Big)} = - {\rm sign}{\Big( B_z \Big)}\,. \label{eq:alfven_sign}
\end{equation}
This equation can be used to detect imprints of torsional Alfvénic pulses in the identified swirls of the stellar models. As we are interested in events that potentially transport energy from the photosphere into the upper atmosphere, our analysis targets the lower and upper photospheric layers of the simulations during the saturation phase of the SSD.

Figure \ref{fig:stats_alfven} shows the bivariate distribution of the average $R_z$ and $R_z^{\rm B}$ for swirls identified in the middle photosphere of the four stellar models during the saturated phase of the SSD. 
The averages are computed over the area of each swirl, and the color indicates the polarity and strength of the vertical component of the magnetic field in which the swirl is embedded.

The pattern revealed by the four plots indicates that the vast majority of swirls obey Eq.\,(\ref{eq:alfven_sign}) in all models, especially in regions where the vertical magnetic field is sufficiently strong. Considering only swirls with an average $|B_z| > 200\,{\rm G}$ over their surface, we find compatibility with Eq.\,(\ref{eq:alfven_sign}) in percentages of $83.1\,\%, 89.3\,\%, 92.1\,\%$, and $90.2\,\%$ for the K8V, K2V, G2V, and F5V models, respectively. Notably, the results for the solar-like model are in agreement with those of \citet{2024A&A...inpress..C}, who performed a similar analysis for a solar atmosphere simulation with an initial vertical magnetic field of $50\,{\rm G}$.

The same analysis is performed for the high photospheric layer of the four stellar models, and the results are shown in Fig.\,\ref{fig:stats_alfven_highphsphere}. The pattern indicating compliance with Eq.\,(\ref{eq:alfven_sign}) 
persists in all four panels of the figure, but a higher level of noise is evident, especially in the K2V model. In this case, the compatibility with Eq.\,(\ref{eq:alfven_sign})
decreases to $69.1\,\%, 58.0\,\%, 66.3\,\%$, and $80.2\,\%$ for the K8V, K2V, G2V, and F5V models, respectively, with a threshold of $|B_z| > 50\,{\rm G}$. This trend mirrors the results obtained by \citet{2024A&A...inpress..C} in simulations of the solar atmosphere. The cause may be a higher proportion of vortices locally generated by shocks that are not associated with torsional Alfvén pulses, or due to the upper boundary conditions of the simulations.


\subsubsection{Energetics}
\label{subsubsec:energetics}
\begin{figure}
	\centering
	\resizebox{\hsize}{!}{\includegraphics{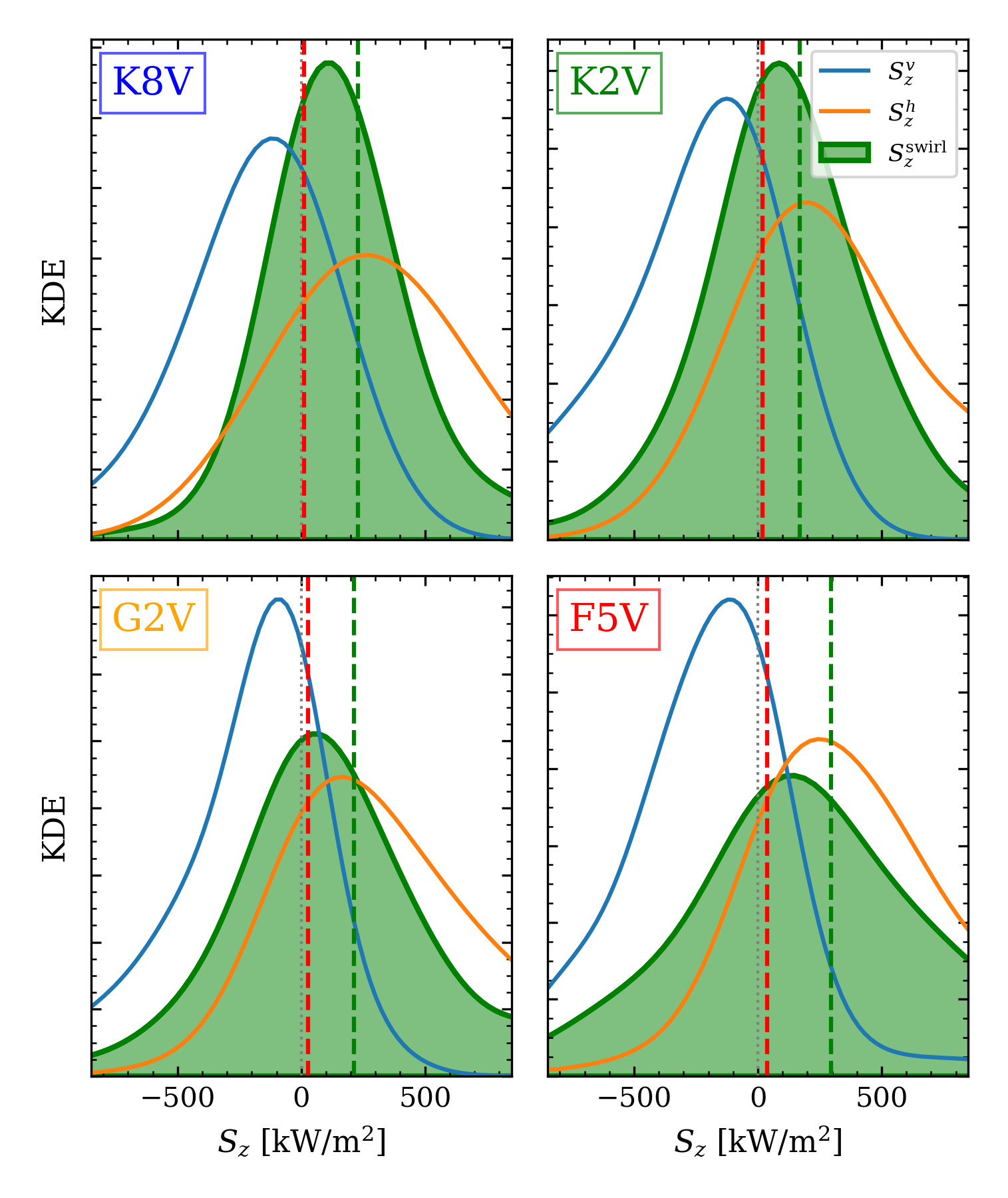}}
	\caption{Kernel density estimation (KDE) of the distribution of the vertical component of the Poynting flux vector associated with Alfvénic swirls, $S_z^{\rm swirl}$ (green). For each entry into the distribution, the corresponding quantity is the mean over the area of those swirls that are  compatible with torsional Alfvénic waves in the middle photosphere of the four stellar models. The KDEs of the horizontal and vertical terms of $S_z^{\rm swirl}$, $S_z^{\rm h}$ and $S_z^{\rm v}$, are shown in orange and blue, respectively. The red vertical dashed lines represent the mean vertical component of the Poynting flux vector computed over the entire domain at the middle photosphere, $S_z^{\rm tot}$, while the green vertical dashed lines represent the mean values of $S_z^{\rm swirl}$.}
	\label{fig:stats_poynting_flux}
\end{figure}
\begin{table}
    \renewcommand\arraystretch{1.2}
    \centering
    \caption{Mean vertical Poynting flux in the middle photosphere computed over the surface of Alfvénic swirls, $S_z^{\rm swirl}$, and over the entire simulation domain, $S_z^{\rm tot}$. See text in Sect.\,\ref{subsubsec:energetics} for more details.}  
    \begin{tabular}{rcccc}
    \hline\hline 
    Poynting flux & K8V & K2V & G2V & F5V \\
    \hline
    $S_z^{\rm swirl}$ [${\rm kW}\,{\rm m}^{-2}$]& 229.6 & 169.1 & 213.3 & 296.2 \\
    $S_z^{\rm tot}$ [${\rm kW}\,{\rm m}^{-2}$]& 11.5 & 18.4 & 26.9 & 36.5 \\
    \hline\\

    \end{tabular}
    \label{tab:poynting_flux}
\end{table}
\begin{table*}
    \renewcommand\arraystretch{1.2}
    \centering
    \caption{Estimated chromospheric flux and basal flux proxy attributed to Alfvénic swirls for the four stellar models. 
    The last row provides the observed lower values of $\log{R^{\prime}_{\rm HK}}$ for stars with similar effective temperatures. See text in Sect.\,\ref{subsubsec:energetics} for more details.}
    \begin{tabular}{rcccc}
    \hline\hline 
     & K8V & K2V & G2V & F5V \\
    \hline
    $F_{\rm swirl}$~~[${\rm erg}\,{\rm cm}^{-2}\,{\rm s}^{-1}$]& $3.5\cdot10^{5}$ & $2.2\cdot10^{5}$ & $4.7\cdot10^{5}$ & $5.5\cdot10^{5}$  \\
    $F_{\rm bol}$~~[${\rm erg}\,{\rm cm}^{-2}\,{\rm s}^{-1}$]& $1.5\cdot10^{10}$ & $3.5\cdot10^{10}$ & $6.3\cdot10^{10}$ & $1.0\cdot10^{11}$ \\
    \hline
    $\log{(F_{\rm swirl}/F_{\rm bol})}$ & -4.6 & -5.2 &-5.1 & -5.3 \\ 
    $\log{R^{\prime}_{\rm HK}}$ & -4.8 & -5.2 & -5.2 & -5.0 \\
    \hline
    \end{tabular}
    \tablefoot{The $\log{R^{\prime}_{\rm HK}}$ values are estimated from \citet[][Fig.\,3]{2018A&A...616A.108B}.}
    \label{tab:basal_flux}
\end{table*}

As we observe a strong correlation between swirls and twist of their magnetic field, indicating compatibility with torsional Alfvénic waves, we investigate whether these events have the potential to transport energy upward in the simulated models. To do this, we use the Poynting flux vector, defined as $\boldsymbol{S}=\boldsymbol{B}\times(\boldsymbol{v}\times\boldsymbol{B})/4\pi$. Since our focus is on the vertical flux, we consider only the vertical component of the Poynting flux vector, $S_z$. Following \citet{2013ApJ...776L...4S}, $S_z$ can be decomposed as
\begin{equation}
S_z = \underbrace{\frac{1}{4\pi} v_z \Big( B_x^2 + B_y^2 \Big) }_{S_z^{\rm v}} - \underbrace{\frac{1}{4\pi} B_z \Big( v_x B_x + v_y B_y \Big)}_{S_z^{\rm h}}, \label{eqn:z-Poynting}
\end{equation}
where $S_z^{\rm v}$ is a term associated with vertical motions of the plasma, while $S_z^{\rm h}$ is associated with horizontal motions of the plasma.

In the following, we restrict our analysis to swirls in the middle photosphere that obey Eq.\,(\ref{eq:alfven_sign}) and are embedded in a sufficiently strong vertical magnetic field ($|B_z| > 200\,{\rm G}$ averaged over the swirl area). 
We will refer to this selection as Alfvénic swirls.

Figure \ref{fig:stats_poynting_flux} shows the distributions of the $S_z^{\rm v}$ and $S_z^{\rm h}$ terms averaged over the area of the Alfvénic swirls in the middle photosphere of the four stellar models. To better visualize the distributions, a kernel density estimation (KDE) is provided. We find that for all four models, the term associated with the vertical motions of the plasma, $S^{\rm v}_z$, (blue curve) peaks at negative values and has a negative skew, while the other term, $S^{\rm h}_z$, (orange curve) is characterized by a positive peak and skew. This behavior is consistent with results from simulations of the solar atmosphere by \citet{2021A&A...649A.121B}.

The sum of the two terms giving the distributions of the vertical component of the Poynting flux vector associated with Alfvénic swirls in the middle photosphere of the four models, $S_z^{\rm swirl}$, is also shown in Fig.\,\ref{fig:stats_poynting_flux} (green curve). These distributions are more centered than the other two, but retain a slight tendency toward positive values, both in terms of peak and skewness. As a result, the means of the four distributions have positive values, as indicated by the green vertical dashed lines in the Fig.\,\ref{fig:stats_poynting_flux} and reported in the first row of Tab.\,\ref{tab:poynting_flux}.

For comparison, in the second row of Tab.\,\ref{tab:poynting_flux} and in Fig.\,\ref{fig:stats_poynting_flux} (vertical red dashed line), we present the average vertical component of the Poynting flux vector computed over the entire domain in the middle photosphere, $S_z^{\rm tot}$. The mean vertical Poynting flux associated with Alfvénic swirls is approximately an order of magnitude larger than the mean vertical Poynting flux in the simulation box at the same height. This implies that Alfvénic swirls are associated with a substantial upward transport of energy relative to the surrounding environment. 
In the following considerations, we assume that only the net vertical Poynting flux of Alfvénic swirls reaches the chromosphere, the rest being reflected or dissipated already in the photosphere.

An important observation is that the mean vertical Poynting flux over the computational domain increases with the effective temperature of the model, while the mean vertical Poynting flux associated with Alfvénic swirls is at a minimum in the K2V stellar model and increases again for model K8V. This result is particularly interesting in the light of studies of the ratio of the chromospheric \ion{Ca}{ii} H and K flux to the bolometric flux, $\log{R^{\prime}_{\rm HK}}$, for main-sequence stellar populations \citep[see, e.g.,][Fig.\,3]{2018A&A...616A.108B}. These studies show that the lower bound of chromospheric emission, known as the basal flux, is minimal and constant for stars within the range $0.5 \leq B-V \leq 1.1$ ($ 6000\,{\rm K} \gtrsim T_{\rm eff} \gtrsim 4500\,{\rm K}$) and increases linearly for cooler stars down to $B-V \sim 1.4$ ($T_{\rm eff} \sim 4000\,{\rm K}$). The reason for the high basal fluxes in such cool stars remains unclear. 

The contribution of Alfvénic swirls to the vertical Poynting flux over the total horizontal domain is,
\begin{equation}
    F_{\rm swirl} = \dfrac{1}{A^{\rm tot}}\sum_{\rm i}{A^{\rm swirl}_i S_{z,\,i}^{\rm swirl}} \,, \label{eq:chromospheric_flux_alfvénic_swirls} 
\end{equation}
where we sum over each Alfvénic swirl contribution to the vertical Poynting flux: $A^{\rm swirl}_i = \pi (r_{\rm eff}^{\,i})^2$ is the effective area of the swirl, and $S_{z,\,i}^{\rm swirl}$ is the vertical Poynting flux averaged over the swirl area. We also divide by the total horizontal area of the simulation domain, $A^{\rm tot} = L_x^2$.

For the bolometric flux, we use Stefan-Boltzmann's law, 
\begin{equation}
    F_{\rm bol} = \sigma T_{\rm eff}^4 \, , \label{eq:stefan_boltzmann}
\end{equation}
where $\sigma = 5.7\cdot10^{-5}\,{\rm erg}\,{\rm cm}^{-2}\,{\rm s}^{-1}\,{\rm K}^{-4}$ is the Stefan–Boltzmann constant and $T_{\rm eff}$ is the effective temperature of the star. 

Assuming that a fraction $r$ of $F_{\rm swirl}$  reaches and dissipates in the low chromosphere where the dominant radiative losses take place in the emission lines of \ion{Ca}{ii} H and K, we can compute its contribution to the chromospheric \ion{Ca}{ii} H and K flux $F_{\rm HK}^{\rm swirl} = r F_{\rm swirl}$.
By taking the logarithm of the ratio between the chromospheric H and K flux due to the Alfvénic swirls and the bolometric flux, $\log{R^{\prime,\,{\rm swirl}}_{\rm HK}} = \log{(F^{\rm swirl}_{\rm HK}/F_{\rm bol})}$, we get an estimate of the basal flux due to Alfvénic swirls \citep[see][]{1979ApJS...41...47L}{}{}. 

In Table \ref{tab:basal_flux}, we present the computed values for $F_{\rm swirl}$, $F_{\rm bol}$, and the basal flux proxy with $r=1$ across the four stellar models. The final row includes approximate lower limits of $\log{R^{\prime}_{\rm HK}}$ for stars characterized by $B-V$ color indices of $1.37$ (K8V), $0.90$ (K2V), $0.65$ (G2V), and $0.47$ (F5V). Notably, the basal flux proxy for the K8V model surpasses those of the other stellar models. Furthermore, an approximately $80\%$ fraction of the flux attributed to Alfvénic swirls would suffice to account for the observed basal flux in stars with $B-V \sim 1.37$. While acknowledging the hypothetical  nature of this result, it introduces a novel and intriguing explanation for the significant basal fluxes observed in stars with $1.1 \leq B-V \leq 1.4$.
Concerning the remaining stellar models, the contribution of Alfvénic swirls may feasibly account for the basal flux in stars resembling the $K2V$ and $G2V$ types. However, such a contribution proves insufficient for the $F5V$ model.


\subsection{Period - Temperature scaling relation}
\label{subsec:temperature_period_relation}

In Sect.\,\ref{subsubsec:swirl_physical_properties} we found that the rotational period of surface vortices increases with the effective temperature of stellar models. At first glance, this result may seem counterintuitive. In order to better understand this phenomenon, we construct a simplified analytical model of a stellar surface vortex. For simplicity, we assume a purely hydrostatic model, but a generalization can be obtained by considering the total pressure $p_{\rm tot} = p_{\rm gas} + p_{\rm mag}$ instead of just the gas pressure.

Let us consider a vertically oriented, axially symmetric vortex in rotational hydrostatic equilibrium in the surface layers of a star. We assume a constant angular velocity $\Omega$ and a uniform plasma density $\rho$. To describe the dynamics of this system, we use Euler's momentum equation,
\begin{equation}
    \rho \frac{D \boldsymbol{v}}{Dt} = -\boldsymbol{\nabla} p + \boldsymbol{F} \,, \label{eq:euler_eq}
\end{equation}
where $D\boldsymbol{v}/Dt$ represents the material derivative applied to the velocity field $\boldsymbol{v}$, $p$ is the gas pressure, and $\boldsymbol{F}$ accounts for external forces. 

\begin{figure}
    \centering
    \resizebox{\hsize}{!}{\includegraphics{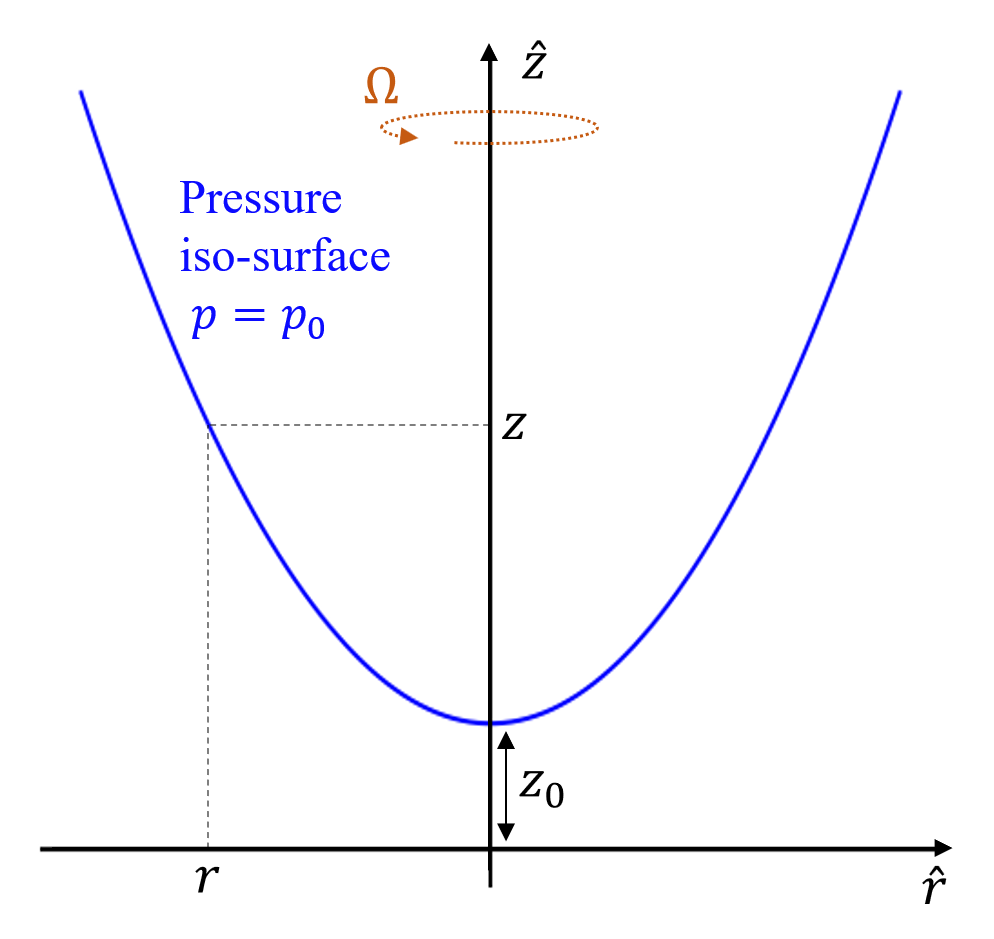}}
    \caption{Two-dimensional representation of an axially symmetric flow rotating at angular velocity $\Omega$. The surface of constant pressure $p = p_0$ is shown, where $p_0$ is the pressure at $(r,z) = (0, z_0)$.}
	\label{fig:model}
\end{figure}

Equilibrium along the radial and vertical directions, $\hat{r}$ and $\hat{z}$, implies that,
\begin{align}
    \frac{\partial p}{\partial r} & = \rho \Omega^2 r \, , \nonumber \\
    \frac{\partial p}{\partial z} & = -\rho g \,. \label{eq:hydro_eq_vortex_cool_stars}
\end{align}
Furthermore, due to axial symmetry, the gas pressure $p$ depends only on the radius $r$ and height $z$. Consequently, we can formulate the total derivative using the following differential expressions,
\begin{equation}
    {\rm d}p = \frac{\partial p}{\partial r} \,{\rm d}r + \frac{\partial p}{\partial z}\, {\rm d}z = \rho \Omega^2 r \, {\rm d}r - \rho g \, {\rm d}z \, , \label{eq:total_derivative_vortex}
\end{equation}
where we have substituted the partial derivatives with the right-hand sides of Eq.\,(\ref{eq:hydro_eq_vortex_cool_stars}). After integration, we obtain a typical pressure equation for a rotating fluid in hydrostatic equilibrium, 
\begin{equation}
    p - p_0 = \frac{1}{2}\rho\Omega^2 r^2 - \rho g (z - z_0) \, , \label{eq:pressure_rotating_fluid}
\end{equation}
where $p$ and $p_0$ are the gas pressure at coordinates $(r,z)$ and $(0, z_0)$, respectively. In particular, Eq.\,(\ref{eq:pressure_rotating_fluid}) indicates that pressure iso-surfaces in vortex flows exhibit a parabolic shape, with the vortex core characterized by lower pressure compared to the periphery. This behavior is illustrated in Fig.\,\ref{fig:model}.

If we consider a pressure isosurface ($p=p_0$), we can derive the following scaling relation,
\begin{equation}
    H_{\rm p} \sim \frac{\Omega^2 r_{\rm eff}^2}{g} \, , \label{eq:H_p_r_relation}
\end{equation}
where the pressure scale height, $H_{\rm p}$, is used as the typical vertical length scale, replacing $z-z_0$, and the radius of the vortex, $r_{\rm eff}$, as the typical radius, $r$. Rearranging Eq.\,(\ref{eq:H_p_r_relation}) and introducing the definition $v_{\phi,\,{\rm eff}} = \Omega r_{\rm eff}$, we obtain,
\begin{equation}
    v_{\phi,\,{\rm eff}}^2 \sim H_{\rm p} g \sim T_{\rm eff} \, , \label{eq:v_T_relation}
\end{equation}
where we exploit the property of a stratified atmosphere where the pressure scale height $H_{\rm p}$ is proportional to $T_{\rm eff}/g$. This relationship allows the model to accurately capture the observed trend shown in the bottom panel of Fig.\,\ref{fig:stats_swirl_properties}. Specifically, the model predicts that the rotational velocity of the vortices increases with the effective temperature of the star.

This scaling relation is consistent with the results of \citet{2016A&A...596A..43C}, who found a similar relationship by considering the pressure, density, and temperature contrasts between the vortex core and the surrounding plasma.

Considering the empirical observation shown in the second panel of Fig.\,\ref{fig:stats_swirl_properties}, which indicates that $r_{\rm eff}/H_{\rm p} \sim {\rm const}$, we can refine Eq.\,(\ref{eq:H_p_r_relation}) as follows,
\begin{equation}
    \frac{1}{\Omega^2} \sim \frac{H_{\rm p}}{g} \, .
\end{equation}
Finally, using the relations $P_{\rm eff} = 2\pi/\Omega$ and $H_{\rm p} \sim T_{\rm eff}/g$, we can deduce that,
\begin{equation}
    P_{\rm eff} \sim \sqrt{\frac{T_{\rm eff}}{g^2}} \, . \label{eq:P_t_g_relation}
\end{equation}
This equation establishes a correlation between the period of rotation of the vortex and key properties of the stellar model, namely $g$ and $T_{\rm eff}$.

To derive a scaling relation for the rotational period that depends only on the effective temperature, we can use the Stefan-Boltzmann law and rely on empirical scaling relations established for cool, main-sequence stars \citep{1991Ap&SS.181..313D},
\begin{align}
L & \sim R^2 T_{\rm eff}^4 \, , \nonumber \\
L & \sim M^{3.7} \, , \nonumber \\
R & \sim M^{0.9} \, , \label{eq:scaling_relations}
\end{align}
where $L$, $M$, and $R$ represent the luminosity, mass, and radius of the star, respectively. When combined with $g \sim M/R^2$ at the star's surface, these relations lead to the following empirical scaling relation,
\begin{equation}
P_{\rm eff} \sim T_{\rm eff}^{2.2} \, . \label{eq:empirical_P_T_relationship}
\end{equation}
This equation confirms that the rotational period for a simplified vortex model increases with the effective temperature of the star, which supports the statistical results obtained from the simulations.
 
%
%

\section{Summary and conclusions}
\label{sec:conclusions}

In this paper, we carried out an analysis on three-dimensional, radiative-MHD, numerical simulations of the atmospheres of main sequence dwarfs of the spectral types K8V, K2V, G2V, and F5V. The focus was on the properties of vortex motions within different layers of the simulation domain. In particular, we investigated how these properties are affected by the strength of the magnetic field, which is amplified by the action of a sub-surface SSD.

As expected, vortex motions are found to be ubiquitous in numerical simulations of stellar atmospheres also other than the solar one. However, we found that different properties of the vortices depend on the characteristics of the stellar model and on the strength of the surface magnetic field. The presence of a strong magnetic field led to a reduction in the number and rotational speed of vortices in the surface layers of the convection zone, due to magnetic quenching. However, in the photosphere, an increased number of vortices showing higher rotational velocities were found in all four stellar models with respect to the kinematic phase, highlighting the fundamental role of the magnetic field in the formation and dynamics of small-scale swirls in stellar atmospheres. 

The size of the vortices was observed to be larger in hotter stars. However, when normalized to the average pressure scale height at the surface, the average radius of the swirls became nearly constant across the different stellar models and layers, being about one fourth of the pressure scale height. This result suggests a direct correlation between the size of the vortices and the scale of the flows, particularly the granular and intergranular flows. However, to firmly establish this correlation, a resolution study is still needed to assess how this result is being affected by the numerical spatial resolution of the simulations \cite[see, e.g.,][]{2020ApJ...894L..17Y}.

We investigate the relation between stellar swirls and perturbations in the surface magnetic fields generated by saturated SSD action. Our investigation reveals a robust correlation between swirls and twists in the magnetic field, suggesting that over $80\,\%$ of the swirls in the middle photosphere are consistent with torsional Alfvénic waves. These events, especially when associated with a sufficiently strong vertical magnetic field, contribute to an average, upward energy flux that significantly exceeds the average Poynting flux over the entire simulation box. This result highlights that swirls associated with Alfvénic pulses, termed Alfvénic swirls, have the potential to play a significant role in the transport of energy to the upper layers of stellar atmospheres.

Moreover, our investigation reveals that the average vertical Poynting flux associated with Alfvénic swirls is higher in the K8V model compared to the K2V model. This observation holds particular significance in the context of studies on chromospheric activity in main sequence stars. Our proxy for the basal flux originating from Alfvénic swirls in diverse stellar models aligns with recent observational data. Specifically, if approximately $80\%$ of the vertical Poynting flux linked to Alfvénic swirls in the K8V model reaches the low chromosphere and dissipates, it could potentially explain the heightened basal flux observed in stars with $B-V \sim 1.4$.

If this vertical Poynting flux is associated with torsional Alfvénic waves propagating along magnetic flux tubes connecting the photosphere and the low chromosphere, the fraction reaching the low chromosphere might be close to unity due to an efficient transport mechanism. Conversely, the majority of the vertical Poynting flux in the photosphere is a result of random motions, with most of it being either reflected or dissipating still in the photosphere \citep[see,][Fig.\,11]{2021A&A...649A.121B}. Therefore, we propose that only a minute fraction of the Poynting flux generated in the photosphere manages to reach the upper layers. Within this small proportion, Alfvénic swirls likely play a predominant role. This hypothesis  introduces a novel framework for interpreting the heightened basal fluxes observed in stars within the range $1.1 \leq B-V \leq 1.4$, offering valuable insights into understanding chromospheric activity in cool stars.

Furthermore, we discovered that the rotational period of swirls increases with the effective temperature of the stellar model. This finding is not trivial, considering that hotter stars typically exhibit faster swirl rotations due to higher flow velocities. Based on a simple model of a vortex in rotational hydrostatic equilibrium and incorporating empirical mass-luminosity and mass-radius scaling relations for cool, main-sequence stars, we derived a scaling relation connecting the rotational period with the effective temperature of the stellar model, $P_{\rm eff} \propto T_{\rm eff}^{2.2}$. This relation supports the results obtained from the statistical analysis.

In conclusion, our study demonstrates the existence of swirls in numerical models of stellar atmospheres, with their specific properties depending on the properties of the stellar models. In addition, we propose that torsional Alfvénic pulses associated with small-scale swirling motions may be an important contributor to the basal flux in \ion{Ca}{ii} H and K (S-index) of cool stars, in particular with respect to the enhanced basal flux in the range $1.1 \leq B-V \leq 1.4$. Although this hypothesis involves some crude assumptions, it effectively reproduces the trend of the observed fluxes and is therefore worth considering.  To make further progress, simulations of a wider parameter space and a finer model grid is needed for a more detailed comparison with measurements of the S-index. Also for these simulations, the chromospheric layers and corresponding physics should be taken into account, which would make it possible to compute the \ion{Ca}{ii} H and K fluxes directly from the models and to compute a synthetic S-index and $R^{\prime}_{\rm HK}$ conversion. Including and excluding the areas of Alfv\'enic swirls would then provide more insights into the role of these kind of swirls for the chromospheric basal flux of cool stars. By continuing to study these phenomena, we can gain a more complete understanding of the intricate processes that take place in stellar atmospheres and their impact on stellar systems.

%
%

\begin{acknowledgements}
The authors acknowledge support by the Swiss National Science Foundation under grant ID 200020\_182094 and the University of Zürich under grant UZH Candoc 2022 ID 7104. This work has profited from discussions with the team of K.\,Tziotiou and E.\,Scullion (conveners) ``The Nature and Physics of Vortex Flows in Solar Plasma'' and with the team of P. Keys (convener) ``WaLSA: Waves in the Lower Solar Atmosphere at High Resolution'' (\url{www.walsa.com}), 
both at the International Space Science Institute (ISSI). Part of the numerical simulations were carried out on Piz Daint at CSCS under project IDs s1059, s1172, and u14, and the rest was carried out on the HPC ICS cluster at USI.
\end{acknowledgements}

%
%


%
%

\bibliographystyle{aa} 
\bibliography{biblio.bib} 

%
%

\begin{appendix}

\section{Surface distribution of swirls: density}
\label{app:surface_distribution_density}
\begin{figure*}[t]
	\centering
	\resizebox{\hsize}{!}{\includegraphics{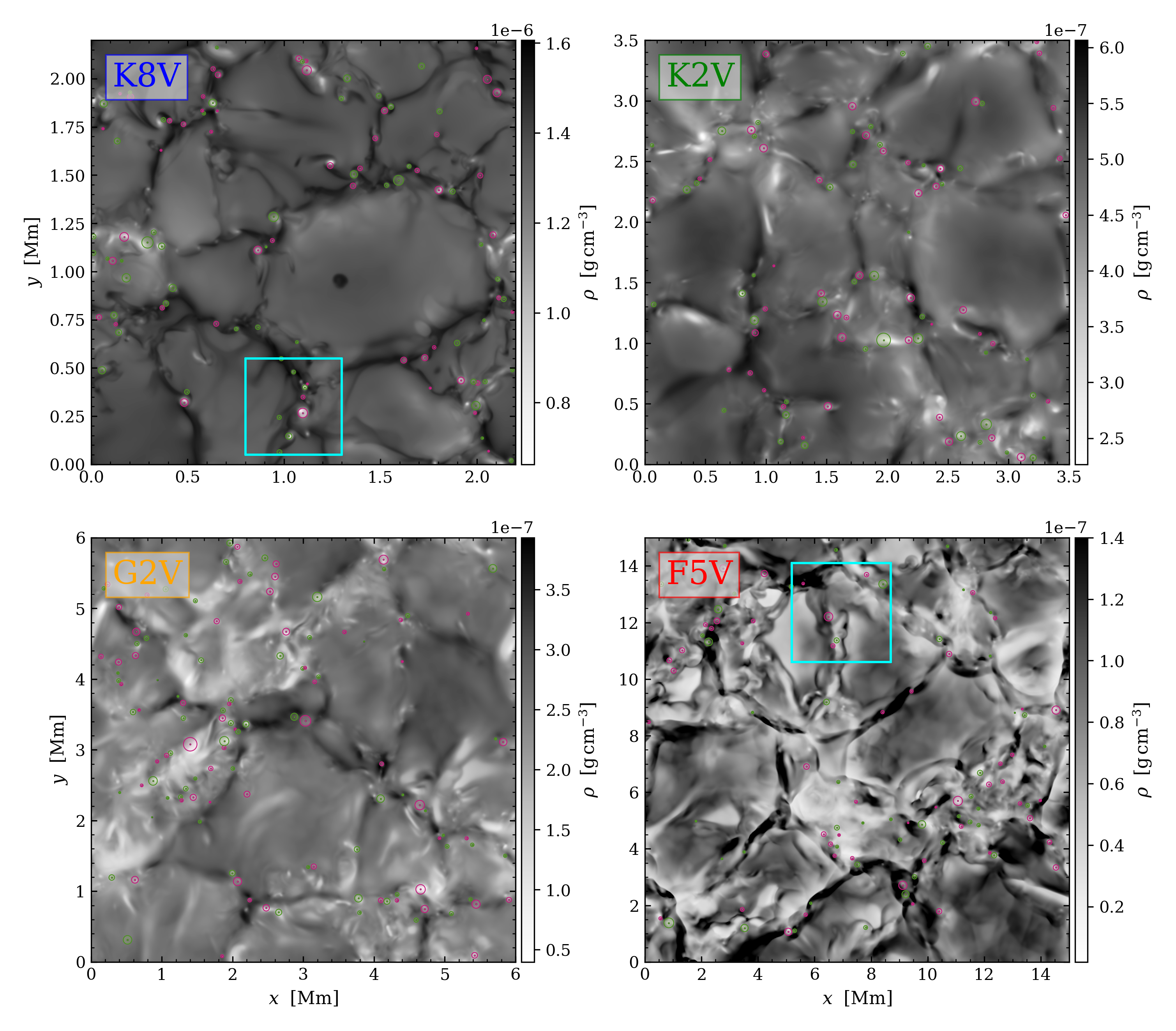}}
	\caption{Density $\rho$ at the average $\tau_{\rm R}=1$ surface for the K8V, K2V, G2V, and F5V models during the kinematic phase. Pink and green disks correspond to the clockwise and counter-clockwise vortices identified by the SWIRL algorithm. The blue squares in the K8V and F5V panels represent the boundaries of the zoom-in plots shown in Fig.\,\ref{fig:visualization_zoomed_rho}.
    }
	\label{fig:visualization_rho}
\end{figure*}
\begin{figure*}
	\centering
	\resizebox{\hsize}{!}{\includegraphics{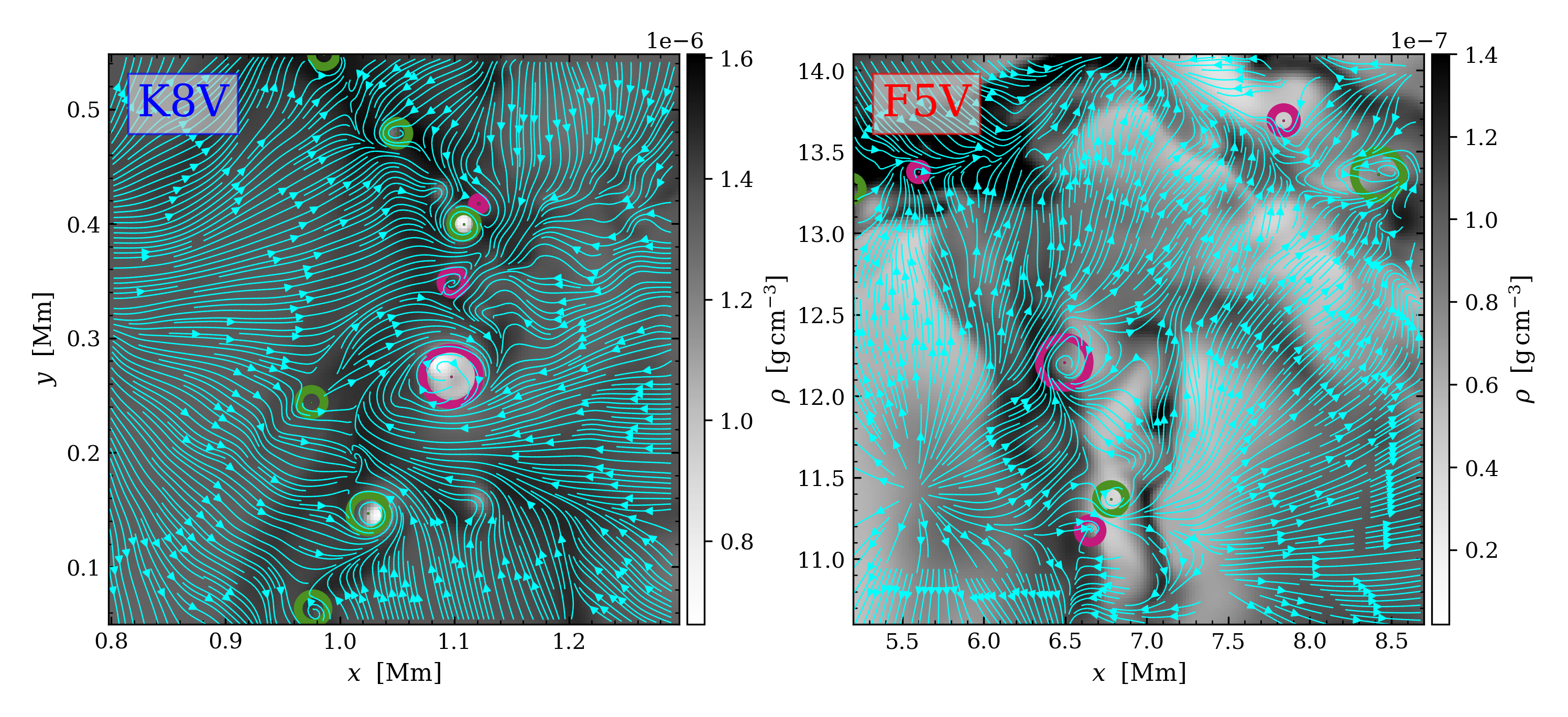}}
	\caption{Identified swirls in the areas depicted in blue in Fig.\,\ref{fig:visualization_rho} for the K8V and F5V models. Pink disks represent clockwise swirls, green counter-clockwise swirls. The horizontal component of the velocity field is represented by blue instantaneous streamlines.}
	\label{fig:visualization_zoomed_rho}
\end{figure*}

Figure \ref{fig:visualization_rho} shows  the density $\rho$ at the average $\tau_{\rm R}=1$ surface for the K8V, K2V, G2V, and F5V models at the same time instances as in Fig.\,\ref{fig:visualization_Ic}. The figure shows a granulation pattern consistent with that of Fig.\,\ref{fig:visualization_Ic}, characterized by cells of bright and warm upflowing plasma surrounded by denser, cooler intergranular downdrafts. Figure \ref{fig:visualization_zoomed_rho} zooms in on two regions of the K8V and F5V models, accompanied by instantaneous streamlines of the horizontal velocity field. Similar to Figs.\,\ref{fig:visualization_Ic} and \ref{fig:visualization_zoomin_Ic}, the vortices identified by the SWIRL algorithm are represented by disks which size and color is defined by their effective radius and direction of rotation, respectively. 

These figures confirm the visual impression conveyed by Fig.\,\ref{fig:visualization_Ic}, indicating that the majority of vortices are located within or in close proximity to intergranular flows. It is also apparent that certain bright features observed at bolometric intensity, commonly referred to as non-magnetic bright spots, are associated with low density regions and swirling motions. A prominent example is shown in the left panel of Figs.\,\ref{fig:visualization_zoomin_Ic} and \ref{fig:visualization_zoomed_rho} at $(x,y) = (1.10,0.25)\,{\rm Mm}$. These events, previously observed in numerical simulations of the solar atmosphere by \citet{2016A&A...596A..43C}, are here identified in simulations of other stellar types as well.

\end{appendix}

%
%
\end{document}